\documentclass[twocolumn,prc,aps,floatfix]{revtex4}
\usepackage[dvips]{graphicx}
\begin{document}
\title{Slow long-range decay of
bound Hartree-Fock orbitals and enhancement of the exchange interaction and tunneling}
\author{V.V. Flambaum}
\affiliation{
 School of Physics, The University of New South Wales, Sydney NSW
2052, Australia
}
\date{\today}
\begin{abstract}
Exchange interaction strongly influences the long-range behaviour of
localised electron orbitals. It   violates the oscillation
 theorem (creates extra nodes) and produces a power-law decay instead of the
usual exponential decrease at large distances. For inner orbitals inside
 molecules decay is $r^{-2}$, for macroscopic systems
 $\cos{(k_f r)} r^{-\nu}$, where $k_f$ is the Fermi momentum and 
   $\nu=3$ for 1D, 3.5 for 2D and 4 for 3D crystal.
 Slow decay increases the exchange interaction
 between localised
 spins and the under-barrier tunneling amplitude.  The
under-barrier transmission
 coefficients in solids (e.g. for point contacts) become temperature-dependent.
\end{abstract}
\maketitle

\section{Introduction}
The Hartree-Fock equation for an electron orbital  $\Psi({\bf r})$ in an atom, molecule or solid has the following form:
\begin{equation}\label{HF}
-\frac{\hbar ^2}{2m}\frac{d^2}{d{\bf r}^2} \Psi({\bf r}) +(U({\bf r})-E) \Psi({\bf r})= K({\bf r})
\end{equation}
\begin{equation}\label{HFK}
 K({\bf r})=\sum_q \Psi_q({\bf r}) \int \Psi_q({\bf r'})^\dagger \frac{e^2}{|{\bf r-r'}|} \Psi({\bf r'}) d {\bf r'}
\end{equation}
Here the summation runs over all electron orbitals  $\Psi_q({\bf r})$ with the same spin projection as  $\Psi({\bf r})$.
Now consider, for example, an inner electron atomic orbital 1$s$. The solution of the Schroedinger equation in 
potential $U({\bf r})$ has  a very small range $a_B/Z$ where $Z$ is the nuclear charge. Outside this range the orbital decreases exponentially
as $\exp{(-r Z/a_b)}$. In the Hartree-Fock equation (\ref{HF}) such rapid decay is impossible.  Indeed, if 
  $ \Psi({\bf r}) \sim \exp{(-r Z/a_b)}$ the left-hand-side of  eq. (\ref{HF}) would be exponentially small while
the right-hand-side is still large since   $K({\bf r})$ in  eq. (\ref{HFK}) contains higher orbitals  $\Psi_q({\bf r})$
which have larger range. The behaviour of the inner Hartree-Fock orbitals inside  atoms have been studied analytically
(in the semiclassical approximation) and numerically in Ref. \cite{DFS}.  The dependence on the radius $r$ can be found from
the multipole expansion of  $|1/({\bf r-r'})|$ in   $K({\bf r})$; the slowest decay normally comes from the dipole
term  $(\sim r'/r^2)$ and/or  last occupied orbital $\Psi_q({\bf r})$, $K \sim \Psi_q({\bf r})/r^2$. The extra nodes
 appear since the orbitals
  $\Psi_q({\bf r})$ oscillate.
For example, the 1$s$ orbital in Cs atom has 3 nodes  \cite{DFS} (without
 the exchange term a ground state has no nodes).
  The existence of extra nodes in solutions of Hartree-Fock equations
was also mentioned in the book  \cite {Froese}.  Outside the atom all orbitals decay with exponential factor for an external
 electron \cite{Handy}.

 Inside solids there are electrons in the conducting band which occupy the whole crystal.
It has been pointed out in Ref. \cite{DFS} that this leads to a long oscillating tail of bound electron orbitals. 
 The effect of the exchange
 interaction $K({\bf r})$ has been estimated in the free band electron
 approximation
  $\Psi_q({\bf r}) = \exp{(i {\bf q \cdot r})}$.
An orbital of a bound electron decreases at large distances as  \cite{DFS}
\begin{equation}\label{tail}
  \Psi({\bf r}) \sim \cos{(k_f r)}/r^4
\end{equation}
 where $k_f$ is the Fermi momentum.

 The derivation of this expression assumes the presence of a partly filled
 conducting electron band. However, in atoms and molecules of any length
the exchange enhancement of the inner orbital tail may be mediated by
 a complete electron shell.
 The question is: can the exchange enhancement in solids be mediated by
a nonconducting electron band? A special interest in this problem may be
 motivated by spintronics and solid state quantum computers based
 on spin qubits.  The long-range tail of the wave function  could,
 in principle,  lead to an enhancement of the exchange spin-spin
 interaction between the distant localised spins, and enhancement of the under-barrier tunneling amplitude. 

  A special feature of the ``long-tail'' mechanism is that
the state of the band electrons does not change, i.e.
  there is no need to have polarization of the conducting band by the localised
spin. The mediating band electrons produce the mean exchange field
 $K({\bf r})$ in  eq. (\ref{HFK}) only.
Therefore,  this ``long-tail'' effect is different from other 
 effects like the RKKI interaction \cite{RKKI} and 
 the double exchange  spin-spin  interaction suggested by Zener \cite{Zener}
 (see also Refs. \cite{Anderson,deGen} and description of Anderson and Kondo
 problems , e.g., in the book \cite{Phillips}).

 To investigate this problem in the present paper we perform calculation
 of the tail using the
Bloch waves and  tight-binding band electron wave functions.  

\section{Atom}

Let us first explain how the long tail appears in atoms \cite{DFS}.
The radial equation for a Hartree-Fock electronic orbital
 $\xi_i(r)=r \phi_i(r)$ is  
\begin{equation}\label{rad}
[-\frac{\hbar^2}{2m}\frac{d^2}{dr^2}+(U_{eff}-E_i)]\xi_i(r)=K_i(r)
\end{equation}
\begin{equation}\label{Ueff}
U_{eff}=U+\frac{\hbar^2l(l+1)}{2mr^2} \ .
\end{equation}
The radial exchange term can be obtained using the multipole expansion of $1/|{\bf r-r'}|$. Outside the radius of an
inner orbital $\xi_i$ (e.g. in the area  $r> a_B/Z$ for $1s$) 
\begin{equation}\label{Ki}
K_i(r)=\sum_{k>0,n} C_{nk} b_{nk}\frac{\xi_n(r)}{r^{k+1}} \ .
\end{equation}
Here $C_{nk}$ are the standard angular momentum dependent coefficients and $b_{nk}=\int r^k \xi_n(r) \xi_i(r) dr$.
For the multipolarity $k=0$ the integral $b_{nk}=0$ due to the orthogonality of radial wave functions
with the same angular momentum. 

  Now we can discuss the large distance behaviour of the orbital $\xi_i(r)$. We will use 1$s$ orbital in  Xe atom ($Z=54$)
 as an example.  The last occupied shells are ...$5s^25p^6$. The orbital $5s$
  does not contribute
to $K_i(r)$ since in this case the multipolarity of the exchange integral is $k=0$ and the orthogonality condition
makes $b_{nk}=0$. The exchange integral $1s5p$ has $k=1$, therefore, at $r \sim a_B$ and outside the atom
$K_{1s}(r)\approx C_{5p,1} b_{5p,1} \xi_{5p}(r)/r^2$.

 The solution of Eq. (\ref{rad}) may be presented as
\begin{equation}\label{xi}
\xi_i(r)=\xi^{free}_i(r)+\xi^{ind}_i(r)
\end{equation}
\begin{equation}\label{xiind}
\xi^{ind}_i(r)=[-\frac{\hbar^2}{2m}\frac{d^2}{dr^2}+(U_{eff}-E_i)]^{-1}K_i(r)
\end{equation}
Outside the radius of the inner orbital ($r>a_B/Z$ for $1s$) the energy $E_i$ is much larger than other
terms in the denominator of Eq. (\ref{xiind}) which are of the order of $E_n$ (since the opertor in the denominator acts
on $\xi_n$). In our example the energy
of $1s$ is $|E_i|=Z^2 \times 13.6$ eV=$4\cdot10^4$ eV while the $5p$ energy is $|E_n| \sim 10$ eV.    
Therefore, we can approximately write 
\begin{equation}\label{xiind1}
\xi^{ind}_i(r)=\frac{K_i(r)}{U_{eff}-E_i} +\frac{\hbar^2}{2m(U_{eff}-E_i)}\frac{d^2}{dr^2}\frac{K_i(r)}{U_{eff}-E_i}+...
\end{equation}
The free solution in this area may be described by the semiclassical (WKB) approximation, 
 \mbox{$\xi^{free}_i(r)\sim |p|^{-1/2}\exp{(-\int |p|dr/\hbar)}$};
 it has the usual range $a_B/Z=0.02 a_B$ for $1s$. Comparison with the numerical
solution of the Hartree-Fock equation for $1s$ orbital has shown that within $\sim$1\%  accuracy
  it is enough to keep the first two terms in the expansion Eq. (\ref{xiind1})  beyond the classical turning point,
  and only one term at $r>10a_B/Z$.
Similar results have been obtained for the Dirac-Hartree-Fock orbitals which include the spin-orbit interaction
and other single-particle relativistic corrections \cite{DFS}. Thus we see that at large distances
\mbox{$\xi_{1s}(r)\approx \rm{const}\, \xi_{5p}(r)/r^2$}.  

\section{1D, 2D and 3D systems}
  If we consider a molecule instead of atom, inner electron orbital will behave
the same way,  \mbox{$\xi_{inner}(r)\approx \rm{const}\, \xi_{valence}(r)/r^2$}.
In macroscopic systems there is a large number of electrons occupying the
valence band and the contribution of different valence electrons interfere
in the exchange term in Eq.(\ref{HFK}). This interference changes the
 long range behaviour.

The equation for a bound electron wave function $\Psi_b({\bf r})$ in a crystal
contains the exchange term from Eq.(\ref{HFK}) describing the exchange interaction of the bound electron
with $2F$  mobile electrons:
\begin{equation}\label{K1}
 K({\bf r})=\int g({\bf r}-{\bf r}')
 [\frac{e^2}{|{\bf r}-{\bf r}'|}-\frac{e^2}{r}] \Psi_b({\bf r}') d {\bf r}',
\end{equation}
\begin{equation}\label{g}
g({\bf r}-{\bf r}')  \equiv \sum_n \Psi_n({\bf r})  \Psi_n({\bf r}')^\dagger .
\end{equation}
Summation goes over F mobile electron  states $\Psi_n({\bf r})$ with the same spin projection.
 To account for the orthogonality condition $\int  \Psi_n({\bf r}')^\dagger\Psi_b({\bf r}') d {\bf r}'=0$
in Eq. (\ref{K1}) we  excluded the zero multipolarity term from the Coulomb integrals, replacing 
$\frac{e^2}{|{\bf r}-{\bf r}'|}$
 by $\frac{e^2}{|{\bf r}-{\bf r}'|}- \frac{e^2}{ r}$.
In the  ``exact'' expression (\ref{K1})  the subtracted term
 $ \frac{e^2}{|{\bf r}|}$ disappears after
 the integration over ${\bf r'}$ since $\int  \Psi_n({\bf r}')^\dagger\Psi_b({\bf r}') d {\bf r}'=0$.

Let us start discussion of crystals
from the simplest problem - a 1D chain of $N$ atoms separated by distance  $a$.
The wave function of a mobile electron can presented as
\begin{equation}\label{mobb}
\Psi_n({\bf r})=L^{-1/2}  e^{ik_n x} v_k({\bf r}),
\end{equation}
where $v_k({\bf r})$ is a periodic function in $x$-direction and $L=Na$ is the length of the chain. 
To perform the summation in Eq. (\ref{g}) analytically we neglect dependence on $k$ in $v_k({\bf r})$.
Taking the standard set of the wave vectors $k_n=2 \pi n/L$, $n=0, \pm 1,...,\pm q$, where $F=2q+1$, we obtain
\begin{equation}\label{gk}
g({\bf r}-{\bf r}')=  v({\bf r})v({\bf r}') \frac{\sin{[k_f (x-x')] }}{x-x'} .
\end{equation}
where $k_F=f \pi/a $ and $f=F/N$ is the band filling factor. Now we can find the exchange term Eq (\ref{K1}).
The leading term in the multipole expansion ($r'<<r$) of \mbox{$\frac{e^2}{|{\bf r}-{\bf r}'|}- \frac{e^2}{ r} \approx\frac{e^2 ({\bf r \cdot r'})}{r^3} $} leads to the
dipole approximation for  $K({\bf r})$ at large distance:
\begin{eqnarray}\label{K2}
\nonumber
 K({\bf r})=\frac{e^2v({\bf r})}{\pi r^3}[\sin{(k_fx)}\int x'\cos{(k_fx')} v({\bf r}')  \Psi_b({\bf r}') d {\bf r}'\\
-\cos{(k_fx)}\int x'\sin{(k_fx')} v({\bf r}')  \Psi_b({\bf r}') d {\bf r}']
\end{eqnarray}

 It is easy to extend the problem
 to 2$D$ and 3$D$ cases. In 2D case we obtain
\begin{equation}\label{gk2}
g({\bf r}-{\bf r}')=  v({\bf r})v({\bf r}') \frac{J_1(k_f R)}{2 \pi R}
\sim \frac{\sin{(k_f R-\pi/4)}} {R^{3/2}} \,
\end{equation}
 where ${\bf R}={\bf r}-{\bf r}'$ and $J_1$ is the Bessel function.
In 3D case
\begin{equation}\label{gk3}
g({\bf r}-{\bf r}')= 
 \frac{ v({\bf r})v({\bf r}') }{2 \pi^2 R^2 }[-\cos{(k_f R)}
+\frac{\sin{(k_f R)}} {k_f R}] .
\end{equation}
Substituting these results into  Eq (\ref{K1}) we obtain in
 the dipole approximation that the exchange
 interaction term decays as
\begin{equation}\label{K3}
 K(r) \sim \cos{(k_f r)} r^{-\nu} \,
\end{equation}
 where $k_f$ is the Fermi momentum and 
   $\nu=3$ for 1D, $\nu=$3.5 for 2D and $\nu=$4 for 3D crystal,
 i.e. $\nu=(5+d)/2$ where $d=1,2,3$ is the dimension.

 Note that the expressions
 (\ref{K2},\ref{K3}) do not vanish if the electron band is complete.
Instead they  have fast oscillations if the electron Fermi momentum  $k_f$ is
 large. This conclusion looks surprising since a complete band does not
 contribute to the  conductivity.  If this conclusion is correct,
one may have an enhanced tunneling amplitude or enhanced exchange interaction
between distant spins (power suppression $r^{-\nu}$ instead of exponential
 suppression) even in non-conducting materials.  However,  this phenomenon may
 be an artefact of the Hartree-Fock approximation for the wave function
(antisymmetric product of the Bloch waves).  We, in fact, have assumed that the
 electrons in complete
band are not localised, i.e. they are still described by the Bloch waves
 (\ref{mobb}) spread over all the crystal. Electron correlations may kill this
 effect, for example, by creating the Mott insulator where all electrons
are localised. On the other hand, it would be incorrect to say that the
 long-range exchange is impossible in principle. For example, correlation
 corrections do not prevent valence electrons from being present on all atoms
 in a molecule where there is no conductivity.

 The long-tail effect  does not appear in any approach where
the exchange interaction is replaced by an effective  potential or by a
density-dependent potential, e.g. in the density functional approach. 
Approximate calculations may also lead to other incorrect conclusions.
For example,  the long-range tail for a complete band case does not appear
  in the tight-binding approximation for the electron wave functions.
 In the tight-binding approximation a wave function of mobile electron is
\begin{equation}\label{mob1}
\Psi_n({\bf r})=N^{-1/2} \sum_l e^{ik_n l a} \Psi_1({\bf r}-l{\bf a}),
\end{equation} 
where $\Psi_1({\bf r}-l{\bf a})$ is the one-site wave function.
 The substitution of $\Psi_n$ from Eq. (\ref{mob1}) into  Eq. (\ref{g})
and summation over $n$ gives the following results:
\begin{equation}\label{sum}
g({\bf r}-{\bf r}')=\sum_{l,m}B(F,l-m)\Psi_1({\bf r}-l{\bf a})
\Psi_1({\bf r'}-m{\bf a}) ^\dagger
\end{equation}
\begin{equation}\label{B}
B(F,l)=\frac{\exp{(i2\pi lF/N)}-1}{N (\exp{(i2\pi l/N)}-1)}\approx \frac{\exp{(i\pi f l)}}{ \pi l}\sin{(\pi fl)},
\end{equation}
where $l>0$, $f=F/N$ is the band filling factor and the last expression is obtained for $l \ll N$. For 
$l=0$ we have $B(F,0)=f$.
Substitution of $g({\bf r}-{\bf r}')$ from Eq. (\ref{sum}) into 
 Eq. (\ref{K1}) shows that  if the band is partly
filled, the tight-binding approximation leads to the same conclusion
$ K(r) \sim \cos{(k_f r)} r^{-\nu}$.
However, for the  completely filled band $f=1$ and $\sin{(\pi fl)}=0$.
This means that the long-range exchange term vanishes in the absence
 of mobile carriers, electrons or holes. The explanation is simple:
 in the tight-binding approximation the
 complete band wave function made of the running waves Eq. (\ref{mob1})
 is equal to the
 antisymmetrised  product of the localised electron wave functions
 $\Psi_1({\bf r}-l{\bf a})$. The exchange interaction with the localised
electrons does not produce the long-range tail. To compare with the Bloch wave
expression one may say that the tight-binding result for the complete
band  corresponds to $K(r) \sim \sin{(k_f r)}=0$ for $r=la$. However,
the  oscillations  of $K(r)$ do not lead to vanishing of its effect on the wave
 functions  - compare with the solution for atomic orbitals in the previous
section.

  At finite temperature  conducting electrons and holes appear. This activates
the long-tail mechanism even in the tight-binding approximation
 and makes the under-barrier transmission coefficient
 temperature dependent. Here it may be appropriate to recall
 that a temperature dependence of the transmission
coefficient has been observed near the  ``0.7 $(2e^2/h)$ structure''
 in the point contact conductance measurements \cite{Thomas96,Thomas}.

This work is  supported by the Australian Research
Council.

\end{document}